\documentclass[3p,times,procedia]{elsarticle}
\usepackage{nupha_ecrc}

\usepackage{subcaption}

\volume{00}

\firstpage{1}

\journalname{Nuclear Physics A}

\runauth{W.A. Zajc}

\jid{nupha}

\jnltitlelogo{Nuclear Physics A}

\usepackage{amssymb}
\usepackage[figuresright]{rotating}

\begin{document}

\begin{frontmatter}



%% Instructions from Editor: Please use the following \dochead only in the preprint version (e-print arXiv etc.); 
%% use empty \dochead{} when submitting to Nuclear Physics A!
\dochead{XXVIth International Conference on Ultrarelativistic Nucleus-Nucleus Collisions\\ (Quark Matter 2017)}
%% e.g. \dochead{17th International Conference on Dynamical Processes in Excited States of Solids}

\title{The Way Forward - Closing Remarks at Quark Matter 2017}


\author{W.A. Zajc}

\address{Physics Department, Columbia University, New York, NY 10027}

\begin{abstract}
This contribution is a written version of my closing talk~\cite{zajcQM17} presented at the Quark Matter 2017~\cite{QM17} conference. Neither the talk nor this contribution to the conference proceedings is intended as a comprehensive summary
\footnote{Those seeking such a summary should consult J\"urgen Schukraft's masterful overview\cite{Jurgen} in the opening talk of the conference.}. 
Rather, a brief discussion is presented of emerging themes and challenges in the field of relativistic heavy ion physics. 
\end{abstract}


\end{frontmatter}

% \linenumbers

%
% WAZ: hack to prevent figure numbering referencing starting at #2 (!)
%
%\setcounter{figure}{0}

\section{Introduction}
The Quark Matter 2017 meeting in Chicago was the latest in a long line of this venerable conference series, which has played a central role in defining the field of relativistic heavy ion physics. More than 200 talks were presented,  the majority of them by young researchers. More than 400 students and postdoctoral fellows attended the pre-conference Student Day~\cite{StudentDay}. It is those researchers who will truly define ``the way forward.'' In what follows I will outline some of the opportunities and challenges that await them. Space restrictions, be they the publisher's page limit or the author's cranial capacity, do not allow for a comprehensive review. The reader is encouraged to prepend to any work cited the phrase ``as but one example."
\label{Sec:Intro}

\section{From Stamp Collecting to Physics}
\label{Sec:Stamps}
The witty and engaging remarks of U.S. Congressional Representative Raja Krishnamoorthi included Ernest Rutherford's famous statement ``All science is either physics or stamp collecting." (While attributed to and characteristic of Rutherford the evidentiary chain is somewhat strained~\cite{StampCollecting}.) But before raising our reductionist flag of victory, we should remind ourselves that our own field has its philatelistic aspects. 
\begin{figure}
  \centering
  \includegraphics[width=0.9\textwidth]{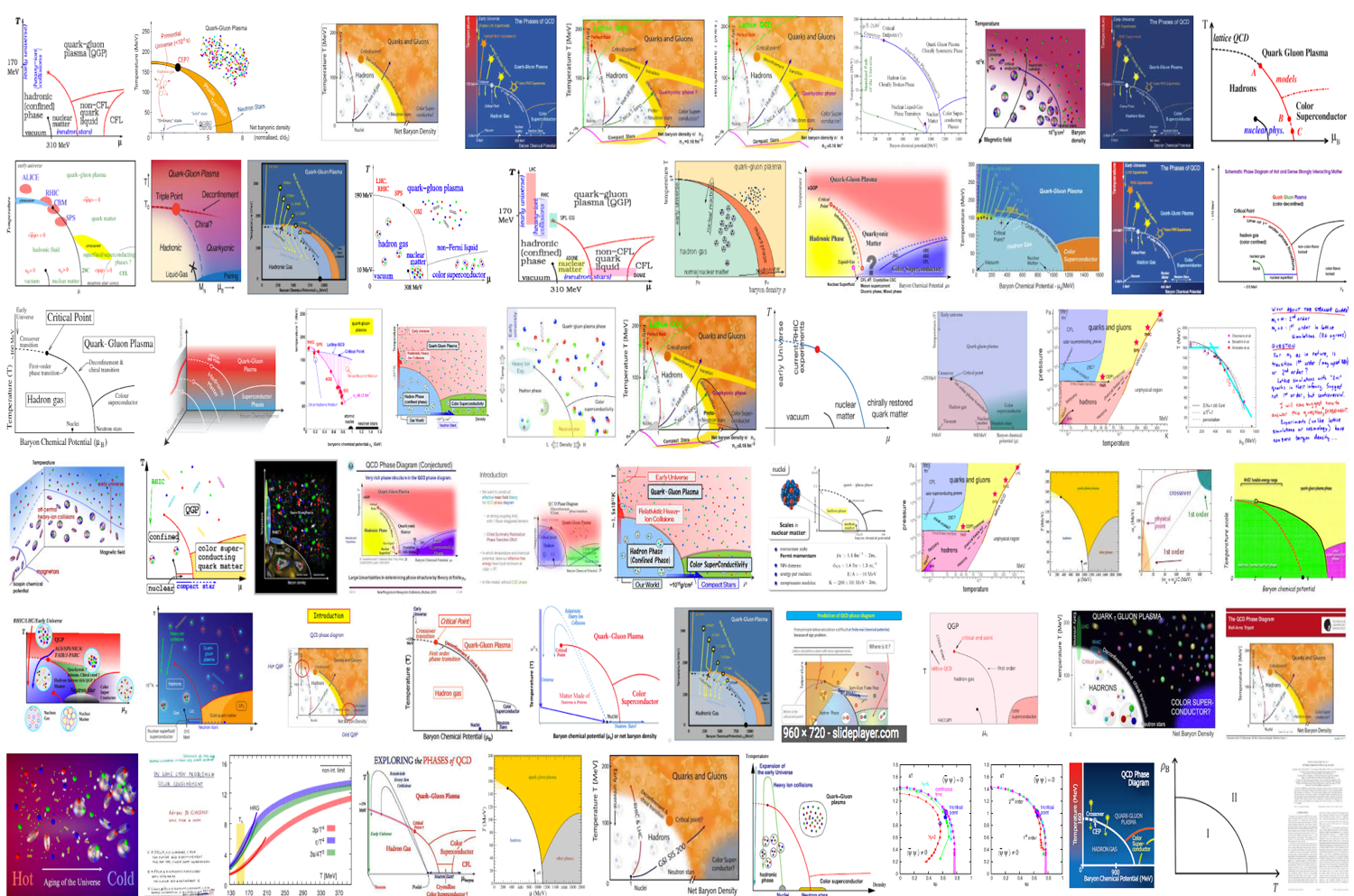}
  \caption{A selection of representations of the QCD phase diagram in the $(\mu_B,T)$ plane.}
   \label{Fig:Stamps}
\end{figure}
That this is the case is suggested by Figure~\ref{Fig:Stamps}, showing the wide variety of schematic representations of the QCD phase diagram. The variety here is not simply reflective of artistic freedom; it is also indicative of scientific ignorance. The incredible successes of the RHIC and LHC heavy ion programs have been narrowly confined to the region $\mu_B \ll T$ where the baryon chemical potential $\mu_B$ is much smaller than the temperature $T$. In this regime the QCD phase transformation is known to be a smooth cross-over, and quantitative contact between experimental observables and the QCD equation of state is being made via lattice QCD (LQCD) calculations and Bayesian methodologies (see Section~\ref{Sec:Precision}). 

It has long been predicted that thermal QCD exhibits a critical end point at finite $\mu_B/T$~\cite{Barducci:1989wi,Halasz:1998qr,Berges:1998rc}. 
Discovery of this feature along with the phenomena expected with a first-order phase transition would greatly expand our knowledge of the QCD phase diagram.
In the next decade experimental measurements will be made in the energy range from $\sqrt{s_{NN}} \sim $~40~GeV down to $\sim$2~GeV required to access the region of the phase diagram where $\mu_B \sim T$. The RHIC Beam Energy Scan~II program\cite{STARoverview} and fixed target measurements at the CERN SPS by NA61/SHINE~\cite{NA61} and at GSI by HADES~\cite{HADES} are underway. There is an associated theory effort that will be essential in interpreting the data and mapping experimental observables to QCD calculations~\cite{Petersen}. 
Both hydrodynamic models and LQCD calculations face challenges in the presence of significant baryon chemical potential, but recent developments on both fronts suggest quantitative precision is within reach. Motivation for this effort is provided not only by the ongoing experimental program but also by the considerable expansion of these investigations that will be enabled by JINR NICA~\cite{NICA}, GSI FAIR~\cite{FAIR}, 
and JPARC-HI~\cite{JPARC}. 
It should also be noted in this context that the recent addition of LHCb to the LHC heavy ion program opens a new window of investigation, with its superb particle identification capabilities at forward rapidities ($\eta  \sim [2,5]$. While at LHC energies one still expects for these larger rapidities  $\mu_B \ll  T$, it is nonetheless a new region in the phase diagram for studying hadrochemistry.

\section{From Discovery to Precision}
\label{Sec:Precision}
More than a decade ago, experimentalists at RHIC issued a call for increased control of correlated errors and systematic uncertainties in theoretical calculations~\cite{RBRCMay04}. Five years later, similar remarks were made in my opening talk at Quark Matter 2009~\cite{Zajc:2009je}. It is enormously gratifying to see 10+ years of sustained theoretical effort come to fruition in recent publications~\cite{Sangaline:2015isa,Pratt:2015zsa,Bernhard:2016tnd} and at this conference~\cite{Bayes}.
\begin{figure}
  \centering
  \includegraphics[width=0.65\textwidth]{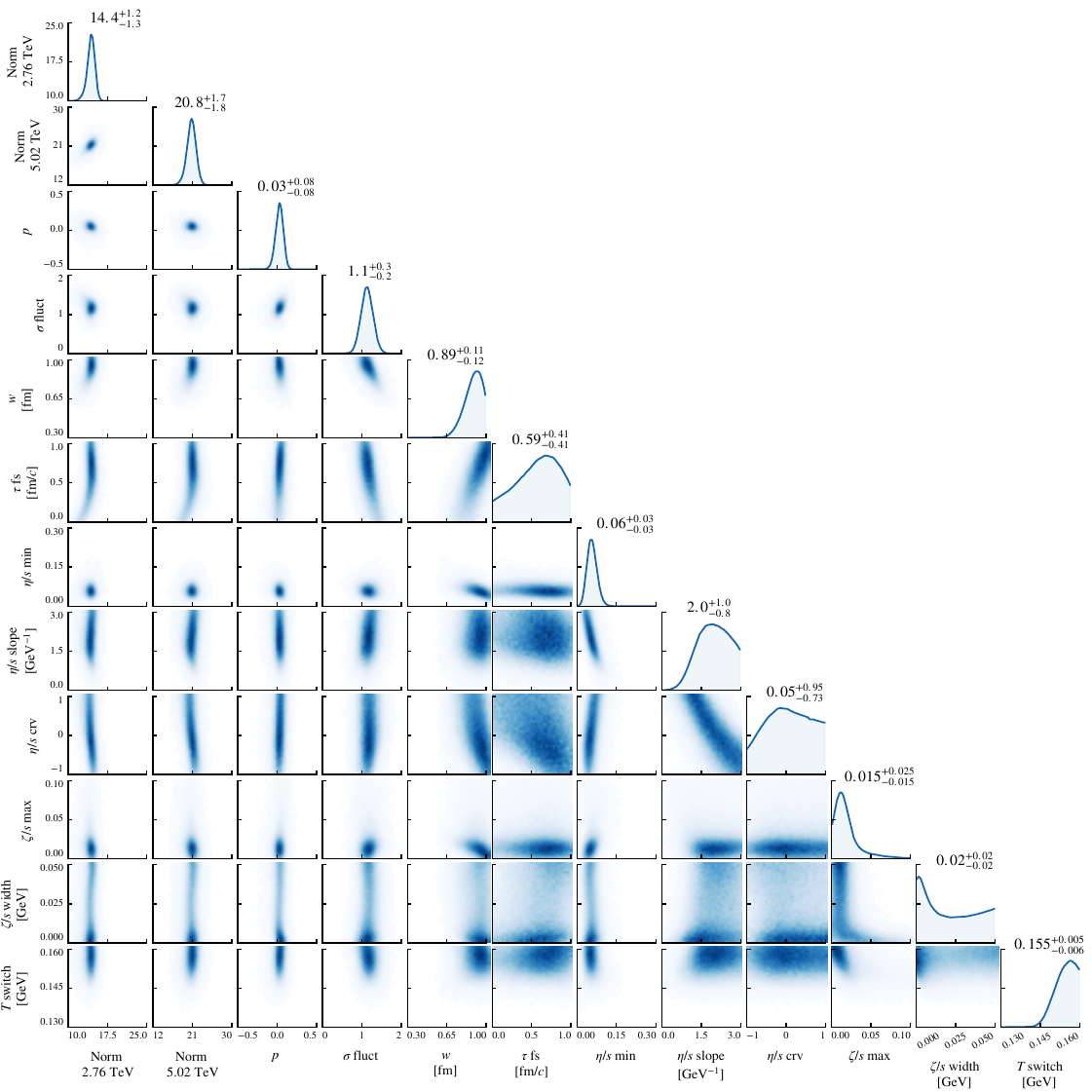}
  \caption{Two dimensional plots of various parameters used in hydrodynamic modeling of LHC collisions as constrained by experimental data~\cite{Bayes}.}
  \label{Fig:Bayes}
\end{figure}
This is illustrated in Figure~\ref{Fig:Bayes}, taken from Ref.~\cite{Bayes}, which shows the range and correlations of various model parameters such as the switching temperature between hydrodynamic and transport calculations, the functional forms of the kinematic shear and bulk viscosities, etc., as constrained by (some of) the experimental data from the LHC at two different energies. Results such as these represent an enormous advance in our ability to quantify key parameters such as the shear viscosity to entropy density ratio $\eta/s$. At the same time, we should not be seduced by the appeal of a literal interpretation of these distributions as the allowed error budget. Systematic uncertainties, both from the experimental data and in the Bayesian emulation process, have yet to be quantified. Nonetheless, a methodology is now in place that did not exist a decade ago, and that methodology is well-suited to the forthcoming era of precision data sets coupled with increasing computational resources. At this conference we saw a particularly intriguing first result using the Bayesian approach to study the dependence of $\eta/s$ on $\sqrt{s_{NN}}$~\cite{Auvinen}; a concerted effort on this front would be very helpful in determining the way forward.

\section{An Essential Conundrum}
\label{Sec:HydroPuzzle}
Central to the precision modeling described in the previous section is the hydrodynamic paradigm, with the majority of the extracted parameters describing initialization of and dissipation in the fluid. The essential role of  hydrodynamics in describing the collisions of large nuclei at relativistic energies is indeed the dominant theme of heavy ion physics in the previous decade. 
\begin{figure}
  \centering
  \includegraphics[width=\textwidth]{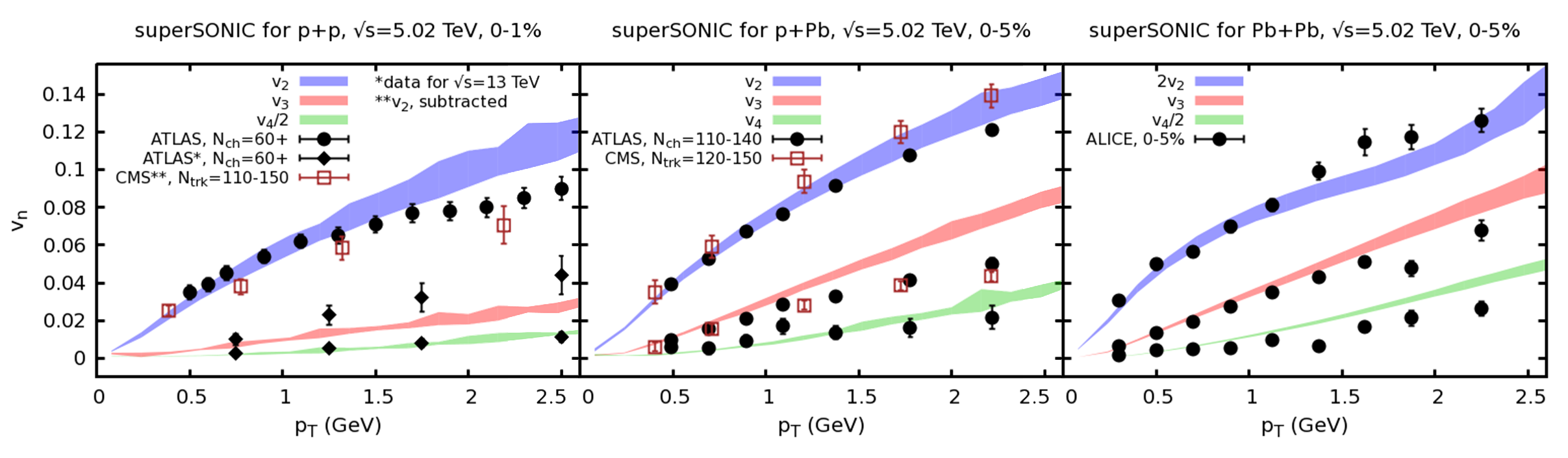}
%  \vspace*{-5cm}
 % \includegraphics[width=0.5\textwidth]{Auvinen.png}
  \caption{Results from hydrodynamic modeling for $v_2$, $v_3$ and $v_4$ versus $p_T$ for p+p, p+Pb and Pb+Pb collisions at 
  $\sqrt{s_{NN}} = 5.02$~TeV~\cite{Weller:2017tsr}. }
  \label{Fig:Paul}
\end{figure}
This decade has seen the discovery of flow-like features in high-multiplicity p+p collisions~\cite{Khachatryan:2010gv} and 
p+Pb collisions at the LHC~\cite{Aad:2013fja,Abelev:2012ola} and d+Au collisions at RHIC~\cite{Adare:2013piz}, leading to the notion of 
``hydrodynamic ubiquity," as embodied in the evocative title of a recent article by Weller and Romatschke 
{\em ``One Fluid to Rule Them All"}~\cite{Weller:2017tsr}.
In this work one unified prescription for initialization of the initial state in p+p, p+A, and A+A collisions is used, along with a common hydrodynamic model, to provide a level of description (Figure~\ref{Fig:Paul}) across these widely disparate systems that would have been regarded as a sensation just a few short years ago. 

In 2017 this result is as likely to be viewed as an irritation as a sensation. Here the irritation is an intellectual one, in that it generates sharp conflicts with other aspects of our understanding. One of those conflicts has been highlighted by Romatschke himself in his talk at this conference, where he demonstrated that ``nuclear collisions are always out of equilibrium"~\cite{PRQM17}, even for large nuclei. While this may be accommodated within a framework that allows ``hydrodynamization"
\footnote{As far as I know, this awkward but accurate construct  appears in the book by Casalderrey-Solana, Liu, Mateos, Rajagopal and Wiedemann  but not in their arXiv article\cite{CasalderreySolana:2011us} with the same title.}
before equiilibration~\cite{casalderrey2014gauge}, it comes with the requirement of strong coupling.
However, the second irritating grain of sand in our evidentiary compilation is the success of models built with the (seemingly) contradictory assumptions of transport of quasi-particles, such as AMPT~\cite{Lin:2001zk}.
\begin{figure}
  \centering
  \includegraphics[width=\textwidth]{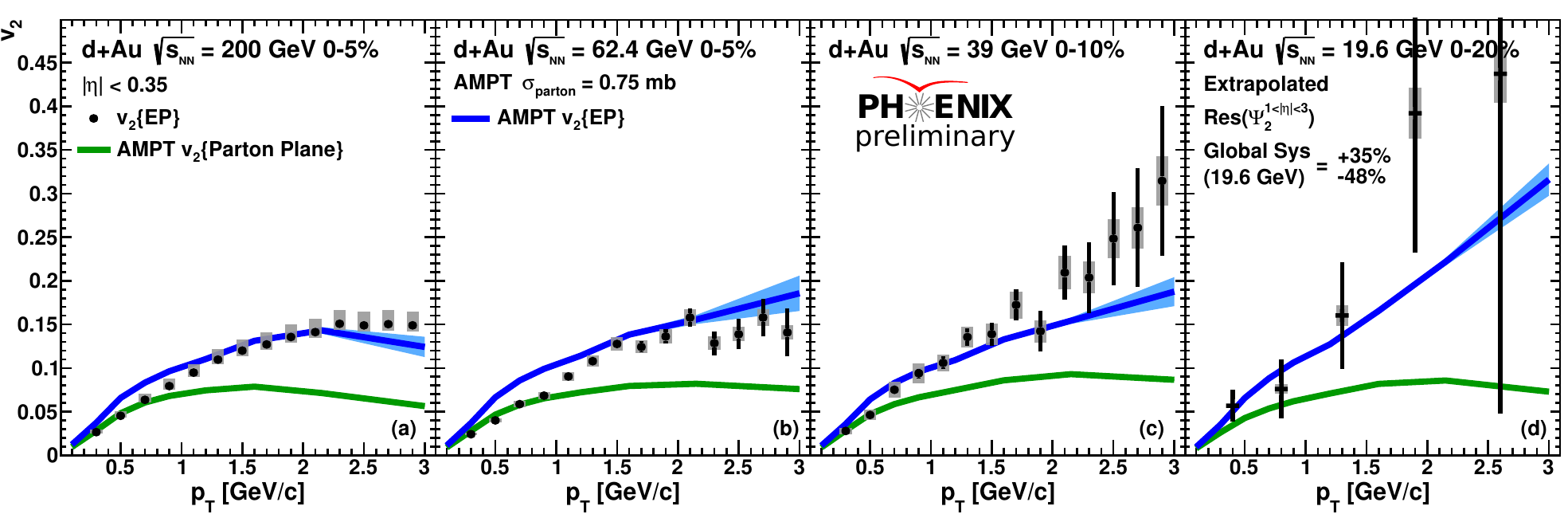}
  \caption{Results from AMPT~\cite{Lin:2001zk} for $v_2$ as a function of $p_T$ compared to PHENIX 
  data for d+Au collisions\cite{JuliaQM17} at a variety of collision energies.}
  \label{Fig:JuliaQM17}
\end{figure}
This is illustrated in Figure~\ref{Fig:JuliaQM17}, which compares PHENIX data for $v_2(p_T)$ in d+Au collisions at collision energies ranging from 
$\sqrt{s_{NN}} = 20$ to 200 GeV. (A similar level of agreement is achieved by AMPT in describing $v_2(p_T)$ at LHC energies~\cite{Bzdak:2014dia}.)

We are thus left with the challenging intellectual puzzle to reconcile two very different and likely incompatible approaches to understanding flow-like phenomena in small systems. This engendered much discussion at the ``Mont Sainte Odile~II" meeting following QM15~\cite{Antinori:2016zxe}. While some progress has been made since that time, particularly in distinguishing the roles played by hydro versus non-hydro modes of the system, it is clear that the process of turning irritating grains of sand to pearls of wisdom remains a slow one - but one that provides beautiful opportunities for new approaches.

\section{A Necessary Dialectic}
\label{CME}
%\begin{figure}
% \centering
%  \includegraphics[width=0.45\textwidth]{STAR_CME}
%  \includegraphics[width=0.50\textwidth]{CMS_CME}
%  \caption{(Left hand side: STAR measurements of the scaled correlator for opposite sign (OS) minus same sign (SS) pairs in p+Au and Au+Au collisions~\cite{Wen}.
%                 Right hand side: CMS measurements of the scaled correlator in p+Pb and Pb+Pb collisions~\cite{Khachatryan:2016got}.}
%  \label{Fig:CME}
%\end{figure}
\begin{figure}
  \centering
  \subcaptionbox{STAR measurements of the scaled correlator for opposite sign (OS) minus same sign (SS) pairs in p+Au and Au+Au collisions~\cite{Wen}.\label{Fig:STAR_CME}}
  {\includegraphics[width=0.45\textwidth]{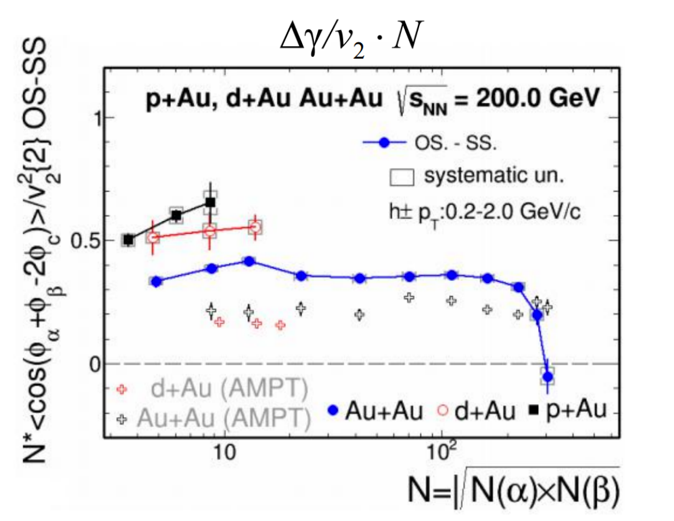}}
  \hfill
  \subcaptionbox{CMS measurements of the scaled correlator in p+Pb and Pb+Pb collisions~\cite{Khachatryan:2016got}.\label{Fig:CMS_CME}}[0.5\textwidth]
  {\includegraphics[width=0.50\textwidth]{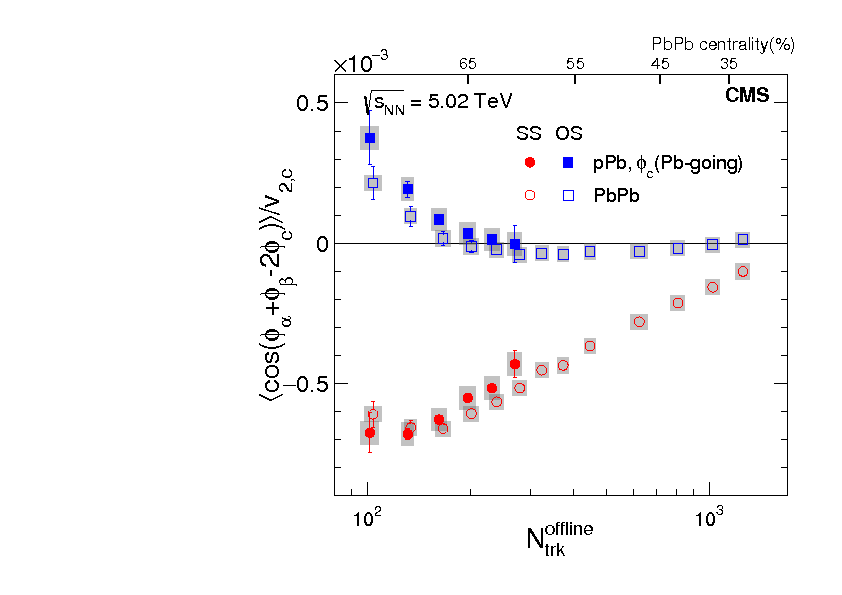}}
  \caption{}
\end{figure}

The past ten years have witnessed a dramatically increased interest in the exploration of the chiral magnetic effect (CME)~\cite{Fukushima:2008xe} and related phenomena~\cite{Kharzeev:2007jp,Kharzeev:2015znc} in relativistic heavy ion collisions. Unambiguous observation of such effects would offer exciting opportunities to explore fundamental topological aspects of gauge theories in a relativistic environment.
Measurements in a non-relativistic condensed matter system have recently been made~\cite{li2016chiral}; that paper has amassed nearly 200 citations in less than two years. At this conference STAR described careful studies~\cite{Wen} using identified particles and comparisons to AMPT to estimate potential backgrounds~\cite{Wen:2016zic}; one such example is shown in Figure~\ref{Fig:STAR_CME}. A new handle on systematics will be provided by an ``isotope run" at RHIC scheduled for 2018, in which ${}^{96}_{44}\mathrm{Ru}$+${}^{96}_{44}\mathrm{Ru}$ collisions will be compared to ${}^{96}_{40}\mathrm{Zr}$+${}^{96}_{40}\mathrm{Zr}$ collisions at the same center-of-mass energy. This will vary the magnetic field created in the collisions by 10\% while controlling for all factors. Needless to say, such a comparison will required control of experimental statistical and systematic {\em relative} uncertainties to well below 10\%. The effect by definition depends on orientation with respect to the reaction plane, so all mundane physical and experimental effects that have a reaction-plane dependence must be carefully identified and eliminated. 

An interesting challenge to CME interpretations was provided by the CMS collaboration with new results from p+Pb collisions. As shown in Figure~\ref{Fig:CMS_CME}, the (scaled) magnitude of the opposite-sign minus like-sign correlator is virtually identical in p+Pb and Pb+Pb collisions. This obviously presents a challenge to any effect dependent on the magnitude of the magnetic field, {\em or} to the scaling method used to compare these two systems. This is precisely the dialectic required to reach firm conclusions in these difficult yet fundamental observations. 

\section{A New Quantitative Tool}
\label{UltraP}
%\begin{figure}
%  \centering
% \includegraphics[width=0.45\textwidth]{S_pb208_2017_v2}
%  \includegraphics[width=0.45\textwidth]{Aaron}
 % \caption{(Left hand side: The nuclear suppression factor versus momentum fraction $x$ calculated~\cite{Guzey:2013xba} from ALICE~\cite{Abelev:2012ba} and CMS~\cite{Khachatryan:2016qhq} measurements of coherent $J/\Psi$ photoproduction in Pb+Pb collisions.
%                 Right hand side: ATLAS measurements~\cite{ATLAS-CONF-2017-011} of the double-differential cross section for photonuclear jet production $d^2\sigma/dH_T dx_A$ versus nuclear momentum fraction $x_A$ for various values of $H_T \approx 2Q$~\cite{Aaron}.}
%  \label{Fig:UltraP}
%\end{figure}
\begin{figure}
  \centering
  \subcaptionbox{The nuclear suppression factor versus momentum fraction $x$ calculated~\cite{Guzey:2013xba} from ALICE~\cite{Abelev:2012ba} and CMS~\cite{Khachatryan:2016qhq} measurements of coherent $J/\Psi$ photoproduction in Pb+Pb collisions.\label{Fig:Guzey}}[0.47\linewidth]
    {\includegraphics[width=0.45\textwidth]{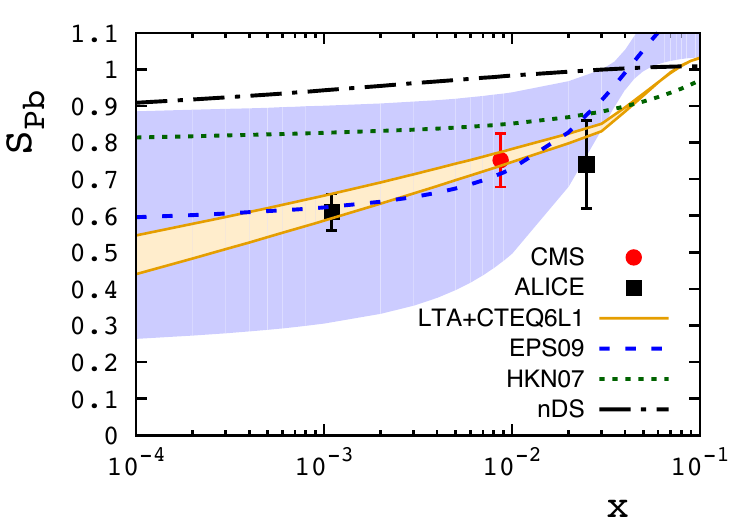}}
  \hfill  
  \subcaptionbox{ATLAS measurements~\cite{ATLAS-CONF-2017-011} of the double-differential cross section for photonuclear jet production $d^2\sigma/dH_T dx_A$ versus nuclear momentum fraction $x_A$ for various values of $H_T \approx 2Q$~\cite{Aaron}.\label{Fig:Aaron}}[0.49\linewidth]  
    {\includegraphics[width=0.40\textwidth]{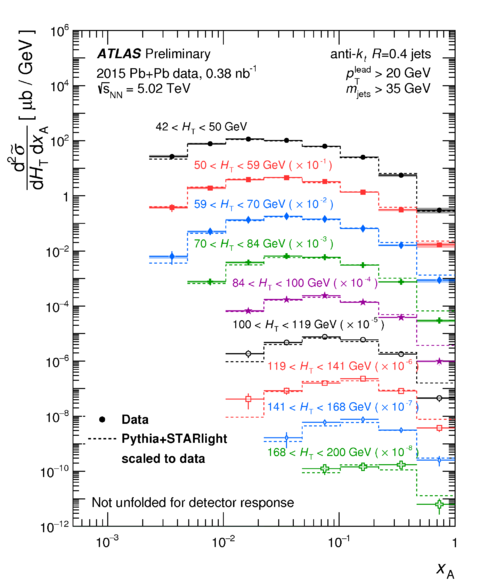}}
    \caption{}
\end{figure}
To date, the LHC has performed p+p, p+Pb and Pb+Pb collisions at collision energies for the nuclear species of $\sqrt{s_{NN}}$ of 2.76 and 5.02~TeV. 
Those collisions involving nuclei also provide a significant flux of virtual photons\cite{Bertulani:2005ru}, making possible an {\em in situ} leptonic deep-inelastic scattering (DIS) program at a hadron facility(!). ALICE~\cite{Abelev:2012ba} and CMS~\cite{Khachatryan:2016qhq} have reported results on coherent $J/\Psi$ photoproduction in ultraperipheral Pb+Pb collisions, which in effect is a measurement of $\gamma^*+\mathrm{Pb} \rightarrow J/\Psi + X$. 
This allows calculation of the nuclear suppression factor for initial state gluons, and results in a very considerable tightening of the error bands allowed by the various structure functions, as shown in Figure~\ref{Fig:Guzey}~\cite{Takaki}. 
At this conference we saw a rather considerable extension from ATLAS~\cite{Aaron} made possible by the analysis of (the much higher cross section) photonuclear jet production. 
These results, shown in Figure~\ref{Fig:Aaron}, are immediately reminiscent of the QCD evolution plots using DIS data that are central to the determination of structure functions~\cite{Aaron:2009aa}. Future work in this vein is likely to provide the best determination of nuclear modifications to structure functions prior to the program at an Electron Ion Collider (see next section).

\section{A Convergence of Interests}
\label{EIC}
As noted in Section~\ref{Sec:HydroPuzzle}, hydrodynamics requires as external input the initial state of the system, where ``initial" in reality means ``some suitable time following the initial collision such that my hydrodynamic model works"
\footnote{This requirement by no means implies "and therefore the assumptions in my hydrodynamic model are justifed".}.
An interesting development at this conference is that the Bayesian methods mentioned previously may be used 
to constrain the fluctuations in the initial state of the {\em proton} in p+Pb collisions~\cite{Moreland}. 

There is of course an entire DIS community dedicated to better understanding the partonic structure of nucleons. The trend in recent years has been to encode that knowledge in generalized parton distributions (GPD's)~\cite{Diehl:2003ny,Belitsky:2005qn}, which provide a relativistic Wigner function formalism to describe parton densities (in one limit) as functions of Bjorken $x$ and the transverse spatial distribution. Measurements to map out the full range of GPD's form a major component of the science that could be performed at a high energy electron ion collider (EIC) with polarized beams. An EIC that can provide $e$+A collisions with heavy nuclei would allow a broad scientific program, greatly increasing the precision of nuclear parton distribution functions and enabling the study of gluon saturation in a truly new regime~\cite{EIC}. The natural evolution of  a component of our community towards such investigations is nicely complemented by the natural evolution of the RHIC facility in the later half of the next decade to host a uniquely capable EIC. 

\section{The Decade(s) to Come}
Our community finds itself in the happy position of having a full set of experimental opportunities at both existing and new facilities. On the first day of the conference Dr.~Tim~Hallman of the U.S.~Department of Energy stated that his office ``is committed to building" sPHENIX,~\cite{sPHENIX} a new experiment with full jet and heavy flavor capability to probe the QGP with the highest resolution possible at RHIC. The ALICE experiment is in the midst of major upgrades~\cite{ALICE_TPC,ALICE_ITS} that will great increase its capabilities in the high luminosity era of the LHC, so that along with ATLAS, CMS and LHCb extraordinary new precision data will become available. As noted in Section~\ref{Sec:Stamps}, over the next decade we can also expect a variety of new facilities and experiments to be constructed to map out the QCD phase diagram in regions of high baryon density. Should this cornucopia be found lacking, there are also new opportunities to be had at a proposed Future Circular Collider~\cite{FCC} with center of mass energy three times that of the LHC. 

This embarrassment of experimental riches will be well-served by a new level of quantitative theoretical understanding made available by increased computational resources, new algorithms, and developing sophistication in techniques and modeling. Our field has risen in stature both by exporting conceptual innovations (e.g., the CME) and importing tools (e.g., AdS/CFT) and deploying them to advance fundamental science. There is every reason to believe the next decade will continue to showcase the essential insights made available through the study of relativistic heavy ion physics.
\label{End}

\section{Acknowledgments}
It is a pleasure to acknowledge contributions and insights provided by Aaron Angerami, Federico Antinori, Steffen Bass, Berndt M\"{u}ller, Brian Cole,  Spencer Klein, Jamie Nagle, J\"{u}rgen Schukraft and Raju Venugopalan, and the support of the U.S. Department of Energy DE-FG02-86ER40281.



%% References


\bibliographystyle{elsarticle-num}
\bibliography{zajcQM17}






% \end{thebibliography}

\end{document}